# Room-temperature relaxor ferroelectricity and photovoltaic effects in SnTiO$_x$/Si thin film heterostructures

Radhe Agarwal[1], Yogesh Sharma[2,3*], Siliang Chang[4], Krishna Pitike[5], Changhee Sohn[3], Serge M. Nakhmanson[5], Christos G. Takoudis[4,6], Ho Nyung Lee[3], James F. Scott[7], Ram S. Katiyar[1], and Seungbum Hong[2,8*]

[1]Department of Physics and Institute for Functional Nanomaterials, University of Puerto Rico, San Juan, PR 00931, USA

[2]Material Science Division, Argonne National Laboratory, Lemont, IL 60439, USA

[3]Oak Ridge National Laboratory, Oak Ridge, Tennessee 37831, United States

[4]Department of Chemical Engineering, University of Illinois at Chicago, Chicago, Illinois 60607, USA

[5]Department of Materials Science and Engineering, Institute of Materials Science, University of Connecticut, Storrs, Connecticut 06269, USA

[6]Department of Bioengineering, University of Illinois at Chicago, Chicago, Illinois 60607, USA

[7]School of Physics and Astronomy, University of St. Andrews, St. Andrews UK

[8]Department of Materials Science and Engineering, KAIST, Daejeon 34141, Republic of Korea

[*]Corresponding authors: sharmay@ornl.gov and seungbum@kaist.ac.kr

# Abstract

We have studied ferroelectricity and photovoltaic effects in atomic layer deposited (ALD) 40-nm thick $SnTiO_x$ films deposited directly onto p-type (001)Si substrate. These films showed well-saturated, square and repeatable hysteresis loops with remnant polarization of 1.5 $\mu C/cm^2$ at room temperature, as detected by out-of-plane polarization versus electric field (P-E) and field cycling measurements. A photo-induced enhancement in ferroelectricity was also observed as the spontaneous polarization increased under white-light illumination. The ferroelectricity exhibits relaxor characteristics with dielectric peak shifting from ca. T = 600K at f = 1 MHz to ca. 500K at 100 Hz. Moreover, our films showed ferroelectric photovoltaic behavior under the illumination of a wide spectrum of light, from visible to ultraviolet regions. A combination of experiment and theoretical calculation provided optical band gap of $SnTiO_x$ films which lies in the visible range of white light spectra. Our study leads a way to develop green ferroelectric $SnTiO_x$ thin films, which are compatible to semiconducting processes, and can be used for various ferroelectric and dielectric applications.

# Introduction

Perovskite type ferroelectric oxides (ABO$_3$) have been studied for decades, because of their versatile multifunctional applications such as sensors, actuators and transducers. Some of the most potential ferroelectric materials include BiFeO$_3$, PbTiO$_3$ or its derivatives such as PbZr$_{1-x}$Ti$_x$O$_3$ [1-5]. However, many environmental issues influenced from toxicity of lead (Pb) have initiated research for environmentally benign (Pb, Bi)-free ferroelectric materials with comparable functionality [6-10]. In the attempts to substitute Pb$^{2+}$ by Ca$^{2+}$ [7, 8] in PbTiO$_3$ (PTO), limited ferroelectricity with lowered tetragonality was observed. Similarly, in other attempts to replace Pb$^{2+}$ with Sr$^{2+}$, strain free SrTiO$_3$ films did not display ferroelectricity at room temperature [9, 10]. However, strained SrTiO$_3$ films showed ferroelectricity at low temperature and the presence of nano-polar regions was detected in both films and bulk single crystals [11, 12], as are domains even in metallic or superconducting SrTiO$_3$ [13]

Recently computational studies suggested substituting Pb$^{2+}$ with isoelectronic Sn$^{2+}$ would give a perovskite structured SnTiO$_3$ with similar polarization as PTO [14-16]. So far attempts to synthesize polar (Sn$^{2+}$)TiO$_3$ by conventional solid state growth route have not been successful [17], because Sn$^{4+}$/Sn$^{2+}$ disproportions led to the formation of nonpolar crystal structure at high temperatures. Although the presence of ferroelectricity has been reported in SnTiO$_x$ films grown by atomic layer deposition (ALD) technique [18] based on well-defined piezoresponse hysteresis loops, there have been no direct evidence via polarization-electric field (P-E) hysteresis loop measurements, no measurement of Curie temperature $T_C$, and no determination of mechanism (displacive, order-disorder, or relaxor).

In addition, it has been predicted by the first principle study that $SnTiO_3$ has an optical band gap of ~2 eV or even lower, which makes it a strong candidate for photo-ferroelectric materials [14, 15, 19]. Photo-ferroelectric materials are particularly interesting due to coupling between photosensitivity and ferroelectric properties, which provide a large possibility of optoelectronics and solar energy harvesting applications [20, 21]. However, there are very few useful photo-ferroelectric materials because most ferroelectrics have wide optical band gaps (> 3 eV).

In this work, we have confirmed ferroelectricity by measuring P-E hysteresis loops, and observed photo-ferroelectricity and photovoltaic effects in non-stoichiometric ($SnTiO_x$) thin films. Furthermore, temperature dependent capacitance and DC leakage current measurements were performed to analyze the relaxor ferroelectricity and conduction mechanisms in these films. Experimental values of the direct band gap in $SnTiO_x$ thin films were also compared with the calculated values obtained by density functional theory (DFT) calculation.

## Results and Discussion

Polarization-electric field (P-E) hysteresis loop measurements were done on capacitors at room temperature in the dark and under white light illumination, as shown in Fig. 1a. The schematic of capacitor $Pt/SnTiO_x/Pt$ structure on p-Si substrate used for ferroelectric and photovoltaic measurements was shown in inset of Fig. 1a. Measurements were done with two top electrodes and a bottom conducting plane of p-Si. P-E hysteresis loops clearly indicate the presence of room temperature ferroelectricity with a well-defined remnant polarization and coercive field. In dark, at ambient temperatures the saturated ($P_s$) and remnant polarization ($P_r$) values were ~3.6 µC/cm² and 1.5 µC/cm², respectively. We also observed P-E hysteresis loops at different poling voltages from 2-8 V and at 100 kHz frequency as shown in Fig. S1a.

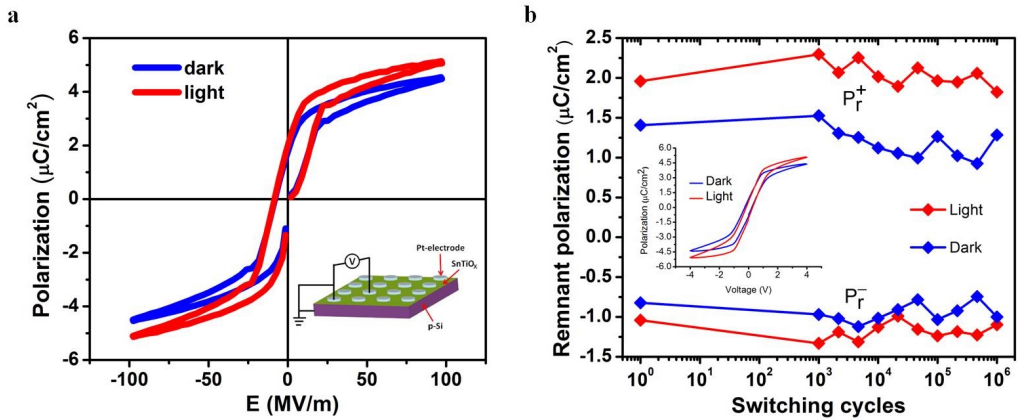

Figure 1. **Room Temperature ferroelectric hysteresis loops. a**, Polarization-electric field (P-E) hysteresis loop at room temperature on Pt/SnTiO$_x$/Pt capacitor in dark and under white-light illumination Inset shows schematic diagram of Pt/SnTiO$_x$/Pt capacitor structure used for electrical measurements. **b**, Evolution of remnant polarization with the number of bipolar cycles. Inset shows the P-E hysteresis loop after $10^6$ switching cycles.

Surprisingly, we found photo-ferroelectric effect in Pt/SnTiO$_x$/Pt capacitor-type structures, which is defined as the photo-induced effect in ferroelectrics that can lead to change in $P_s$ and $P_r$. Under white light illumination, $P_s$ and $P_r$ increased to 4.3 and 1.8 µC/cm², respectively (see Fig. 1a). Such enhanced ferroelectric behavior defines the photo-ferroelectric nature of SnTiO$_x$ thin films. Moreover, polarization fatigue, which is an important factor to determine the stability of switchable polarization over the field cycles, was also measured to evaluate the reliability of these ferroelectric thin films. Figure 1b shows the evolution of remnant polarization of $SnTiO_x$ as a function of switching cycles in dark and under white light illumination. We confirmed nearly fatigue-resistant characteristic up to $10^6$ cycles in our $SnTiO_x$ capacitor, which is comparable to early studies of $Pt/Pb(Zr,Ti)O_3/Pt$ capacitors. Moreover, the ferroelectric nature of these films was found to be stable over wide temperature range from 85 K to 390 K, as shown in Fig. S1b.

Temperature dependent current (I) versus voltage (V) (I-V) curves were measured from 83 K to 600 K to understand the electrical conduction mechanism in $Pt/SnTiO_x/Pt$ capacitors, as shown in Fig. 2a. To understand the dominant conduction mechanism of charge transport, the temperature dependent I-V curves were replotted at logarithmic scale, as shown in Fig. 2b-c. As can be seen from Fig. 2c, the logarithmic I-V plot at room temperature can be fitted into three regimes, where the low and intermediate regimes correspond to Ohmic conduction and Child's square law, respectively, and the high voltage region corresponds to steep increase in current [22]. Based on such I-V behavior, the conduction mechanism in our films can be attributed to trap-assisted space charge limited current conduction [23]. However, it has been observed in oxide thin films that linear I-V relation at low applied voltages can also be explained by Simmons' modified Schottky emission mechanism, which required film-thickness dependent current measurements for further clarification [24]. Furthermore, the temperature dependence of current can be used to determine the depth of the trapping levels. The slope of ln I versus 1/T, which is $(E_c-E_t)/k$, provides the activation energy value as shown in Fig. 2d. We found that shallow trap levels exist near the conduction band with activation energy of 1.3±0.1 eV. This value is reasonable for oxygen-vacancy transport in perovskite oxides, including $SrTiO_3$, where 1.1 eV is most common value [25, 26]. This value is found in two ways: From the I(1/T) graph in Fig. 2d and, less precisely, as the kink and steep increase in current of the ln I versus ln V curve in Fig. 2c near ln V = 0 (i.e., V = ca. 1 V).

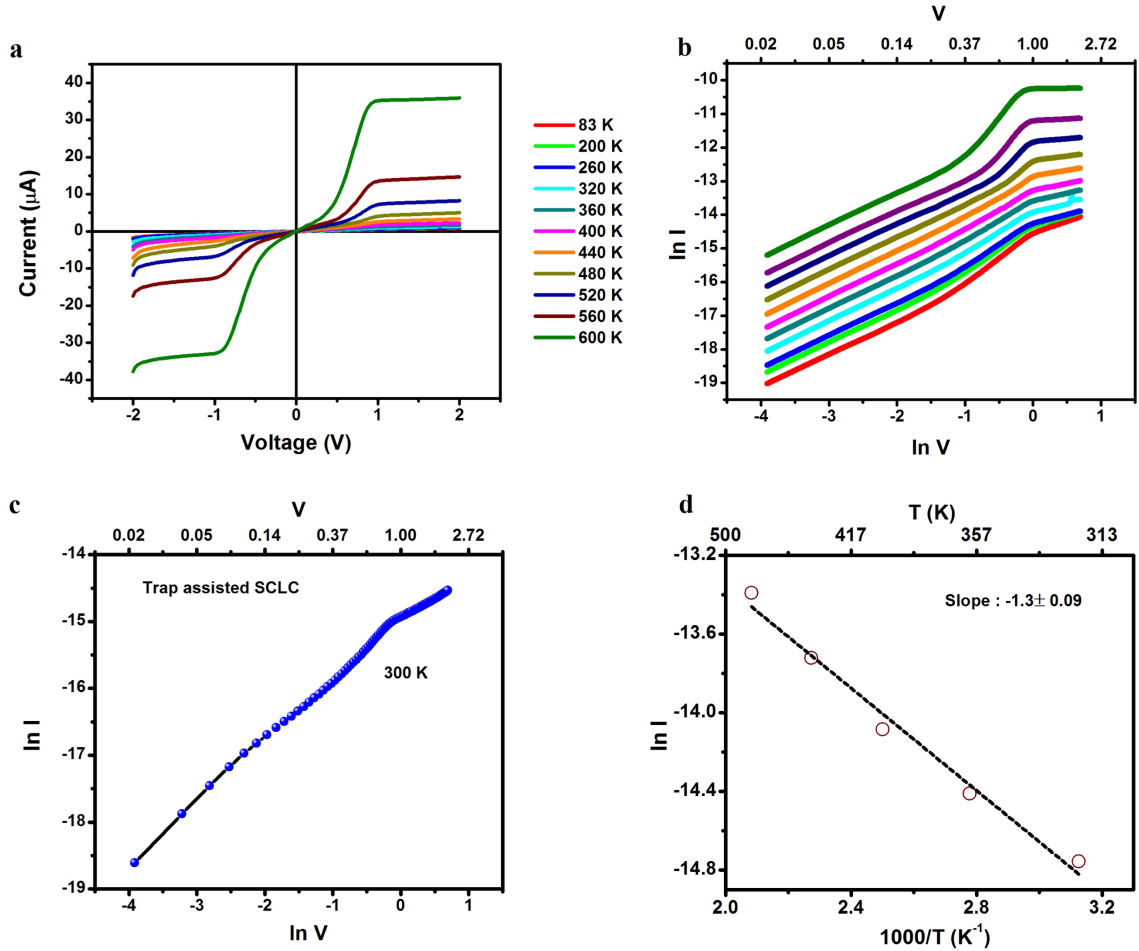

Figure 2. **Temperature dependence of electrical transport measurements.** **a**, I-V characteristics of Pt/SnTiO$_x$/Pt capacitors at various temperatures. **b**, ln I-ln V at various temperatures. **c**, ln I-ln V plot at 300 K exhibits trap assisted space charge limited current (SCLC) behavior of current transportation. **d**, Arrhenius plot for capacitor structure.

It is important in this work to rule out charge injection as a source of artifacts. In addition to noting that the polarization hysteresis loops are saturated, we provide in Fig. 3 data showing a dielectric peak at $T_C$ = 450-600 K, depending upon probe frequency from ca. 500 Hz to 300 kHz. These data display the signature of a relaxor ferroelectric but do not satisfy a Vogel-Fulcher relationship with an extrapolated freezing temperature of ca. 452 K (Fig. 3b), suggesting local disordering of the Sn-ions. A plausible hypothesis for the observed glassy behavior is the presence of some $Sn^{+4}$ at the $Ti^{+4}$ perovskite B-site. This could be tested via XPS measurement of the $Sn^{+4}/Sn^{+2}$ ratio. Since all these temperatures are well above ambient, these properties do not detract from the potential for commercial devices.

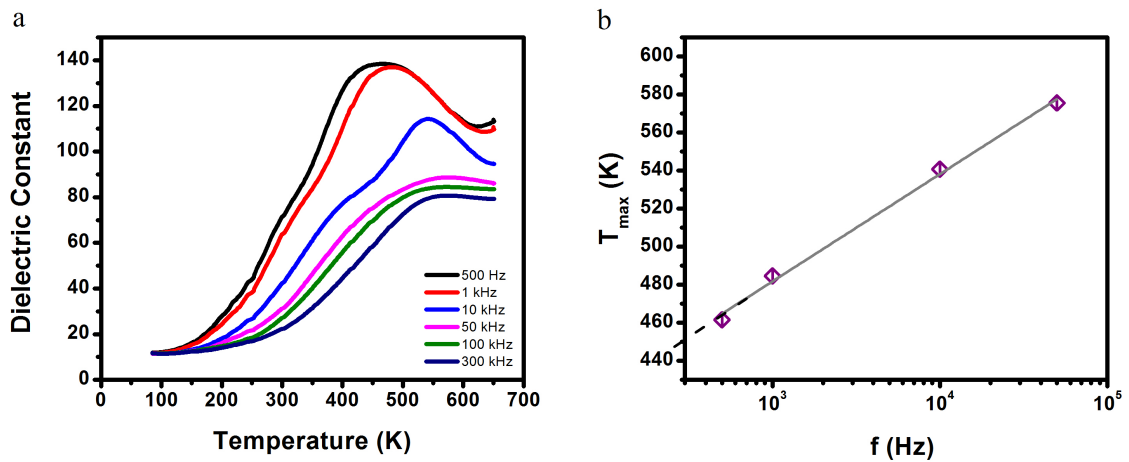

Figure 3. **a**, Temperature dependence of dielectric constant. **b**, Extrapolated freezing temperature was observed to be 452 K.

In addition to the photo-induced changes in $P_s$ and $P_r$, we found a separate photovoltaic effect in as-grown $Pt/SnTiO_x/Pt$ capacitors. In the dark condition, a very small current ($10^{-4}$ μA) at zero bias voltage could be detected. Under white light illumination a distinct photovoltaic behavior was observed with short circuit current ($J_{SC}$) = 3 μA and open circuit voltage ($V_{OC}$) =0.13 V, as shown in Fig. 4a. $V_{OC}$ and $J_{SC}$ are measured as a function of time over multiple on/off light cycles, showing good retention over time with instability in $V_{OC}$ and $J_{SC}$ displaying a sudden increase and then come back to initial values, as shown in Fig. 4b. To explain the unstable nature of $J_{SC}$ in multiple light on/off cycles, we measured variation in $J_{SC}$ under continuous light illumination of 120 seconds, as shown in Fig. S2. We observed a continuous increase in photocurrent under continuous light illumination, which could be attributed to the presence of trap levels due to structural defects, mainly oxygen vacancies [27, 28]. Under the light illumination, photo-generated electrons are transferred to the conduction band and some of them occupy shallow trap levels. Photocurrent starts increasing slowly as the trap levels are filled gradually. In addition, we have also observed photosensitivity of our samples in ultraviolet (UV) region. The observed photovoltaic behavior of $SnTiO_x$ films under UV-light illumination has been shown in Fig. S3.

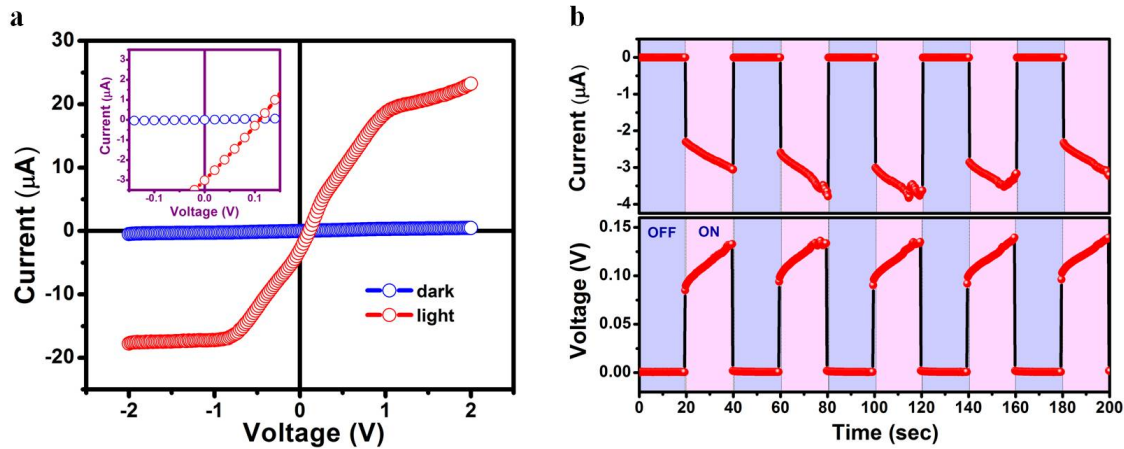

Figure 4. **Photovoltaic effect in Pt/SnTiO$_x$/Pt capacitor. a**, I-V characteristics measured on Pt/SnTiO$_x$/Pt capacitor in dark and under white light illumination. Inset shows a zoom of the I-V curves around zero field, revealing $J_{SC}$ and $V_{OC}$ to be 3 μA and 0.13 V, respectively. **b**, Time dependence of $J_{SC}$ and $V_{OC}$ measured under multiple on/off light cycles.

We performed a combined study of experiment and theoretical calculations to estimate the optical band gap of $SnTiO_x$ films. We directly obtained refractive indices of films, $n$ and $k$, using spectroscopic ellipsometry technique and calculated absorption coefficient, $\alpha$. The direct band gap was determined to be 2.6 eV by linearly extrapolating $(\alpha E)^2$ versus photon energy (E), as indicated by a gray line in Fig. 5. This result is in a good agreement with the first principle study that predicted an optical band gap of ~2 eV, which indeed makes $SnTiO_x$ a strong candidate for photo-ferroelectric materials as confirmed by our current study [14, 15, 19]. However, at this point, the nature of the band gap is not clear as the experiment points toward a direct band gap while the theory suggests an indirect band gap.

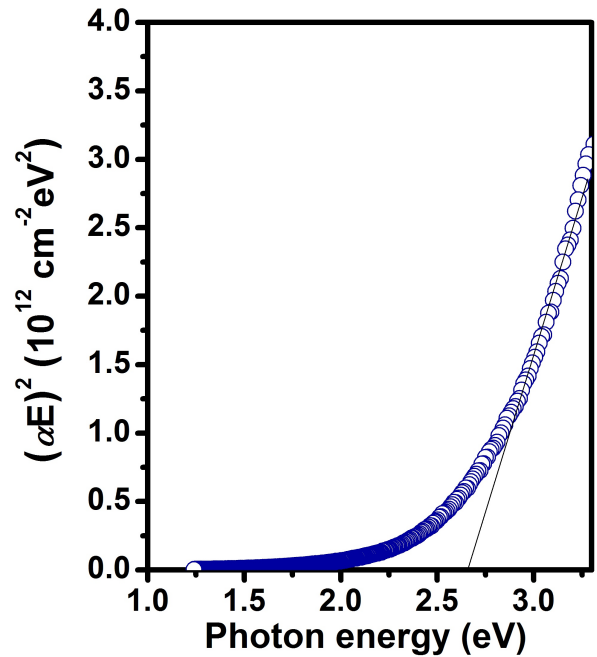

Figure 5. **Optical band gap estimation in $SnTiO_x$ thin films.** $(\alpha E)^2$ versus photon energy plot gives the direct band gap of $SnTiO_x$ thin films grown on p-Si substrate. A gray line is an extrapolation line to estimate the band gap.

Electronic band structure and electronic density of states (EDOS) were investigated to further analyze strain dependent band gap tuning in SnTiO$_3$. Figure 6 shows the electronic bands dispersion diagram (left panel) and electronic density of states (right panel) of SnTiO$_3$ at (a) $\varepsilon = -0.33\%$, (b) $\varepsilon = 0\%$ and (c) $\varepsilon = +0.33\%$. We see that the band gap is indirect from X to Γ across the phase transition from $\varepsilon = -0.33\%$ to $\varepsilon = +0.33\%$. We have also plotted the ion and $l$ quantum-number-resolved EDOS for the aforementioned three epitaxial strain states. The valence band between 16 and 17 eV in each case is composed of Sn $s$ and O $p$ states. The conduction band between 19 and 21 eV is largely composed of Ti $d$ states with a small mixture of O $p$ states. The plot of strain dependent band gap based on EDOS suggests that we need to enhance tensile strain inside SnTiOx film to further lower the optical band gap, which will enhance the photoferroelectric property of the film (see Figure S4).

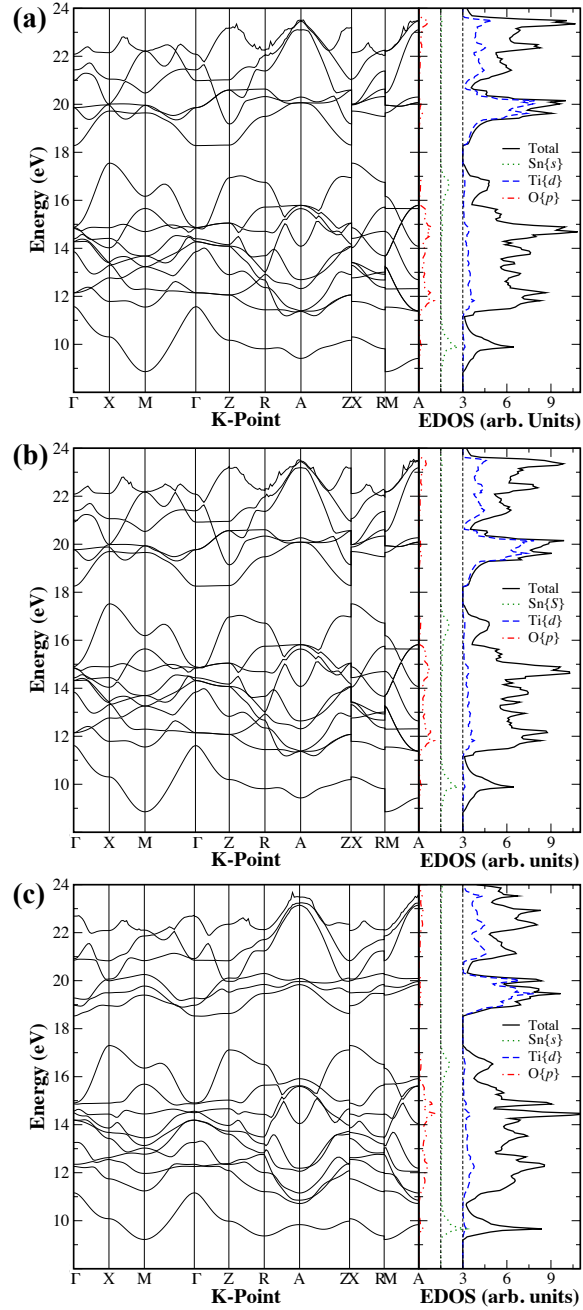

Figure 6. Electronic Bands Dispersion (left panel) and ion and $l$ quantum-number-resolved electronic density of states (EDOS) (right panel) calculated for SnTiO$_3$ for (a) $\varepsilon = -0.33\%$, (b) $\varepsilon = 0\%$ and (c) $\varepsilon = +0.33\%$.

## Conclusion

We demonstrated room temperature ferroelectricity and photovoltaic effects in $SnTiO_x$ thin films deposited on p-type Si substrates. Our films showed well-defined polarization hysteresis loops at room temperature with a relaxor-type transition to paraelectric phase above 500 K. A photo-induced enhancement in spontaneous polarization indicated the photoferroelectric nature of $SnTiO_x$ films. The optical band gap was found to be around 2.6 eV which can be tuned further via strain imposed by substrate. This study provides a path to develop green material for various ferroelectric and dielectric based emerging nanoelectronic devices.

## Methods

**Experimental details.** $SnTiO_x$ thin films of 40 nm thickness were deposited on p-Si substrates using atomic layer deposition (ALD) technique as detailed in ref. 18. Pt-top electrodes were deposited on ALD grown $SnTiO_x$ thin films using rf-magnetron sputtering technique. Macroscopic ferroelectric properties were characterized using RT6000 loop tester (Radiant Technologies). To examine the intrinsic fatigue behavior, 10 μs wide electric pulses with a frequency of 100 Hz were applied to the Pt-top electrode. Photovoltaic measurements were performed using Keithley-2401 electrometer under 1 sun AM 1.5 solar simulator with light source density ~ $1kW/m^2$. Conduction mechanisms in $Pt/SnTiO_x/Pt$ capacitor were studied through current-voltage (I-V) measurements using a Keithley 2401 source-meter unit. A programmable Joule-Thompson thermal stage system (MMR model # K-20) was used for temperature dependent measurements. Dielectric measurements between frequencies 500 Hz and 300 kHz were done using an impedance analyzer HP4924A. M-2000 ellipsometer (J. A. Woollam Co.) was employed to record two ellipsometric parameters, $\psi$ and $\delta$, of SnTiO$_x$ film on p-type Si substrate and the bare substrate. Then, a two-layer model composed of the substrate and film was used to determine the optical constants of the films. The optical constants of the substrate as well as film thickness were obtained separately and fixed during the fitting procedure.

**Theoretical method.** A computational study was conducted to compute the bandgaps of the polar epitaxial phases of pure SnTiO$_3$. DFT [30, 31] calculations were performed using the Vienna Ab initio Simulation Package (VASP) [32, 33] within the local density approximation (LDA), parameterized by Perdew and Zunger [34]. The projector-augmented plane-wave method [35, 36] simulated the core and valence electrons. A plane wave basis supporting the wave function cutoff at 900 eV converged the Cubic $Pm\bar{3}m$, Tetragonal $P4mm$, and Monoclinic $Cm$ Total energy to within 1 meV/f.u.. Zone-edge-shifted 8×8×8 Monkhorst-Pack (MP) [37] k-point mesh, was used for the Brillouin zone (BZ) integration. Force components of individual ions were relaxed to less than 1 meV/Å and the pressure on the simulation cell to less than 0.1 kbar in all studied polymorphs. Monoclinic angle ($\beta$) is not allowed to relax in monoclinic $Cm$ phase. The biaxial misfit strain was defined as $\varepsilon = \frac{a}{a_0} - 1$, where $a_0$ corresponds to the optimized lattice parameter of Cubic $Pm\bar{3}m$ structure with all the normal stresses relaxed to values less than 0.1 kbar. It is a well known that LDA exchange correlation functional underestimates the value of band gap. We have overcome this issue by computing the band gap using the Heyd-Scuseria-Ernzerhof (HSE) screened hybrid functional [14, 38].

**Acknowledgments**

R. A. acknowledges receiving graduate fellowships from NSF-IFN Grant # 1002410. Work at Argonne (Y. S. and S. H., design of experiment, data analysis and contribution to manuscript writing) and ORNL (C.S. and H.N.L., optical spectroscopy) was supported by the U.S. Department of Energy (DOE), Office of Science, Office of Basic Energy Sciences (BES), Materials Sciences and Engineering Division. S.C. is grateful for financial support by the NSF (CDMR #1309114), and C.G.T. acknowledges support by the NSF (CDMR #1309114, CBET #1067424 and EEC #1062943).

# Supporting Information for

# Room-temperature relaxor ferroelectricity and photovoltaic effects in SnTiO$_x$/Si thin film heterostructures

Radhe Agarwal[1], Yogesh Sharma[2,3*], Siliang Chang[4], Krishna Pitike[5], Changhee Sohn[3], Serge M. Nakhmanson[5], Christos G. Takoudis[4,6], Ho Nyung Lee[3], James F. Scott[7], Ram S. Katiyar[1], and Seungbum Hong[2,8*]

[1]Department of Physics and Institute for Functional Nanomaterials, University of Puerto Rico, San Juan, PR 00931, USA

[2]Material Science Division, Argonne National Laboratory, Lemont, IL 60439, USA

[3]Oak Ridge National Laboratory, Oak Ridge, Tennessee 37831, United States

[4]Department of Chemical Engineering, University of Illinois at Chicago, Chicago, Illinois 60607, USA

[5]Department of Materials Science and Engineering, Institute of Materials Science, University of Connecticut, Storrs, Connecticut 06269, USA

[6]Department of Bioengineering, University of Illinois at Chicago, Chicago, Illinois 60607, USA

[7]School of Physics and Astronomy, University of St. Andrews, St. Andrews, UK

[8]Department of Materials Science and Engineering, KAIST, Daejeon 34141, Republic of Korea

*Corresponding authors: sharmay@ornl.gov and seungbum@kaist.ac.kr

**Temperature dependent P-E hysteresis loop mesaurements:**

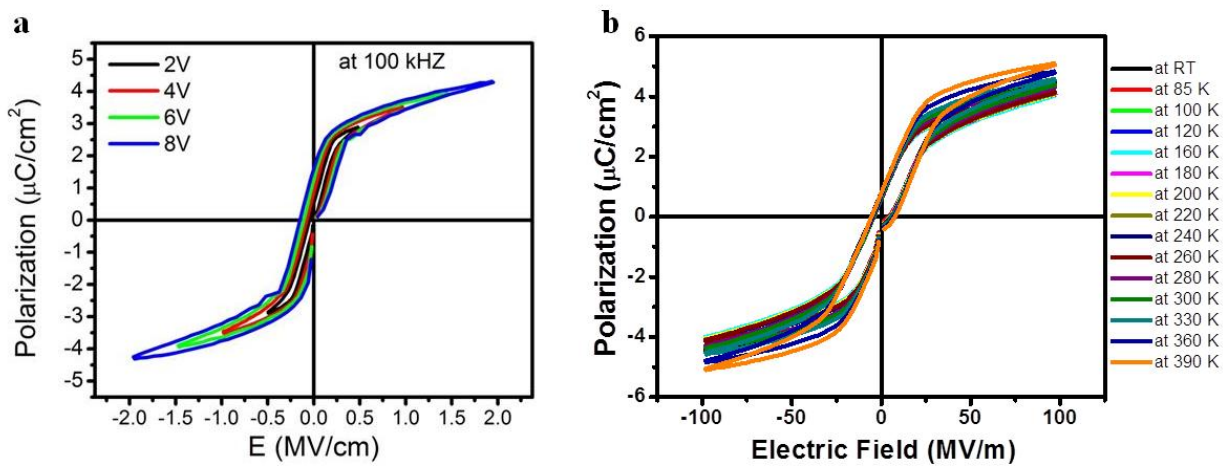

Figure S1. (a) P-E hysteresis loop at different voltages (b) Temperature dependence of P-E hysteresis loop.

**Photocurrent transient measurement:**

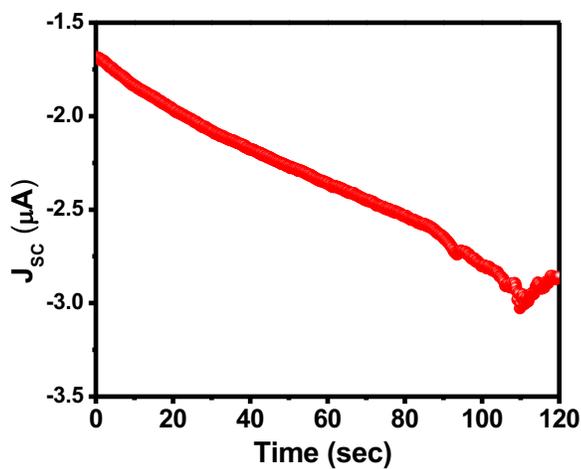

Figure S2. Variation in $J_{SC}$ under continuous white light illumination of 120 seconds.

## UV-assisted photovoltaic measurements:

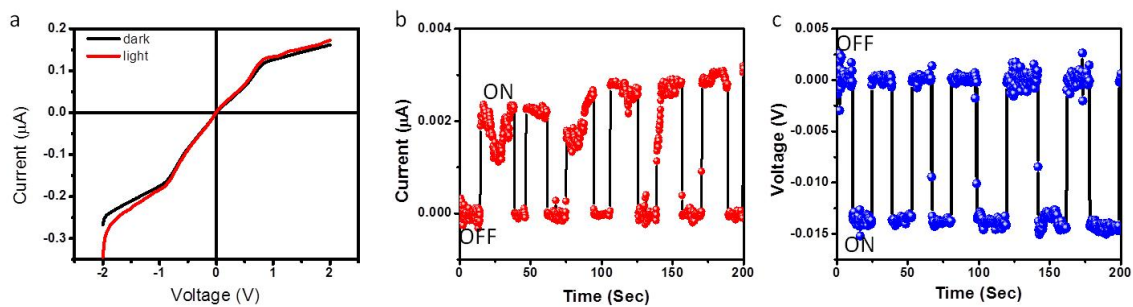

Figure S3. (a) I -V characteristics measured on Pt/SnTiOx/Pt capacitor in dark and under UV light illumination. (b)-(c) Time dependence of $J_{SC}$ and $V_{OC}$ measured under multiple on/off light cycles.

## Strain dependence of optical band gap:

To further understand the dependence of optical band gap on the substrate, strain dependence of electronic properties of $SnTiO_3$ was studied using DFT calculations. Figure 6 shows the energy levels of Valence-band maximum (VBM) and Conduction-band minimum (CBM) computed for epitaxially strained $SnTiO_3$ with the local-density approximation (LDA) and Heyd-Scuseria-Ernzerhof (HSE) exchange correlation functionals represented in dashed and solid lines respectively. The data is shown for $\varepsilon$ ranging within $\pm 2$ (negative and positive values represent compression and tension, respectively). We see that the band-gap increases from 2.175 eV, at $\varepsilon = 0$ to 2.655 eV, at $\varepsilon = +1\%$. We see an abrupt change in band gap value from $\varepsilon = 0\%$ to $\varepsilon = +0.33\%$, as the space group of the structure changes from *P4mm* to *Cm*. The mechanism of the band gap opening was attributed to the hybridization between Ti $d_{xy}$ and O $P_x$ orbitals at Γ point, when the four-fold symmetry is broken from *P4mm* to *Cm* phase [29].

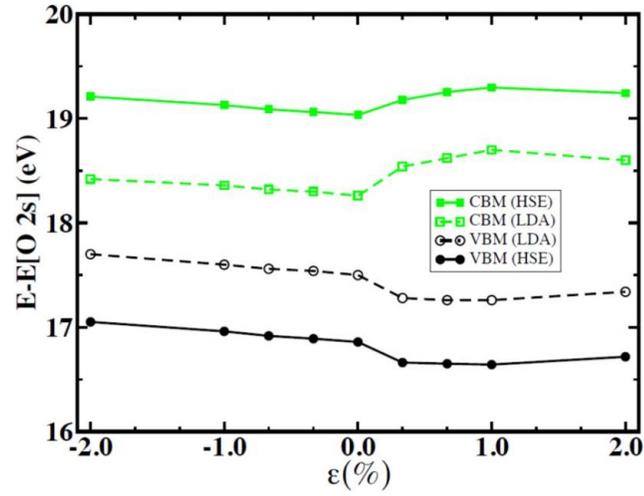

Figure S4. Calculated CBM and VBM using LDA and HSE in polar tetragonal *P4mm* phase represents a band gap of 2.175 eV which increases up to 2.655 eV in misfit strain stabilized polar monoclinic *Cm* phase.